\documentclass[twoside,twocolumn,9pt]{article}
\usepackage{extsizes}
\usepackage[super,sort&compress,comma]{natbib} 
\usepackage[version=3]{mhchem}
\usepackage[left=1.5cm, right=1.5cm, top=1.785cm, bottom=2.0cm]{geometry}
\usepackage{balance}
\usepackage{mathptmx}
\usepackage{amssymb}   % \gtrsim için gerekli
\usepackage{sectsty}
\usepackage{graphicx} 
\usepackage{lastpage}
\usepackage[format=plain,justification=justified,singlelinecheck=false,font={stretch=1.125,small,sf},labelfont=bf,labelsep=space]{caption}
\usepackage{float}
\usepackage{fancyhdr}
\usepackage{fnpos}
\usepackage[english]{babel}
\addto{\captionsenglish}{%
  
}
\usepackage{array}
\usepackage{droidsans}
\usepackage{charter}
\usepackage[T1]{fontenc}
\usepackage[usenames,dvipsnames]{xcolor}
\usepackage{setspace}
\usepackage[compact]{titlesec}
\usepackage{hyperref}
\usepackage{lipsum}  
\usepackage{float}
\usepackage{amssymb}
\usepackage{amsbsy}

\definecolor{cream}{RGB}{222,217,201}

\begin{document}

\pagestyle{fancy}
\thispagestyle{plain}
\fancypagestyle{plain}{
%%%HEADER%%%
\renewcommand{\headrulewidth}{0pt}
}
%%%END OF HEADER%%%

%%%PAGE SETUP - Please do not change any commands within this section%%%
\makeFNbottom
\makeatletter
\renewcommand\LARGE{\@setfontsize\LARGE{15pt}{17}}
\renewcommand\Large{\@setfontsize\Large{12pt}{14}}
\renewcommand\large{\@setfontsize\large{10pt}{12}}
\renewcommand\footnotesize{\@setfontsize\footnotesize{7pt}{10}}
\makeatother

\renewcommand{\thefootnote}{\fnsymbol{footnote}}
\renewcommand\footnoterule{\vspace*{1pt}% 
\color{cream}\hrule width 3.5in height 0.4pt \color{black}\vspace*{5pt}} 
\setcounter{secnumdepth}{5}

\makeatletter 
\renewcommand\@biblabel[1]{#1}            
\renewcommand\@makefntext[1]% 
{\noindent\makebox[0pt][r]{\@thefnmark\,}#1}
\makeatother 
\renewcommand{\figurename}{\small{Fig.}~}
\sectionfont{\sffamily\Large}
\subsectionfont{\normalsize}
\subsubsectionfont{\bf}
\setstretch{1.125} %In particular, please do not alter this line.
\setlength{\skip\footins}{0.8cm}
\setlength{\footnotesep}{0.25cm}
\setlength{\jot}{10pt}
\titlespacing*{\section}{0pt}{4pt}{4pt}
\titlespacing*{\subsection}{0pt}{15pt}{1pt}
%%%END OF PAGE SETUP%%%

%%%FOOTER%%%
\fancyfoot{}
\fancyfoot[LO,RE]{\vspace{-7.1pt}\includegraphics[height=9pt]{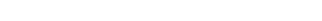}}
\fancyfoot[CO]{\vspace{-7.1pt}\hspace{13.2cm}\includegraphics{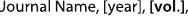}}
\fancyfoot[CE]{\vspace{-7.2pt}\hspace{-14.2cm}\includegraphics{head_foot/RF}}
\fancyfoot[RO]{\footnotesize{\sffamily{1--\pageref{LastPage} ~\textbar  \hspace{2pt}\thepage}}}
\fancyfoot[LE]{\footnotesize{\sffamily{\thepage~\textbar\hspace{3.45cm} 1--\pageref{LastPage}}}}
\fancyhead{}
\renewcommand{\headrulewidth}{0pt} 
\renewcommand{\footrulewidth}{0pt}
\setlength{\arrayrulewidth}{1pt}
\setlength{\columnsep}{6.5mm}
\setlength\bibsep{1pt}
%%%END OF FOOTER%%%

%%%FIGURE SETUP - please do not change any commands within this section%%%
\makeatletter 
\newlength{\figrulesep} 
\setlength{\figrulesep}{0.5\textfloatsep} 

\newcommand{\topfigrule}{\vspace*{-1pt}% 
\noindent{\color{cream}\rule[-\figrulesep]{\columnwidth}{1.5pt}} }

\newcommand{\botfigrule}{\vspace*{-2pt}% 
\noindent{\color{cream}\rule[\figrulesep]{\columnwidth}{1.5pt}} }

\newcommand{\dblfigrule}{\vspace*{-1pt}% 
\noindent{\color{cream}\rule[-\figrulesep]{\textwidth}{1.5pt}} }

\makeatother
%%%END OF FIGURE SETUP%%%

%%%TITLE, AUTHORS AND ABSTRACT%%%
\twocolumn[
  \begin{@twocolumnfalse}
{\includegraphics[height=30pt]{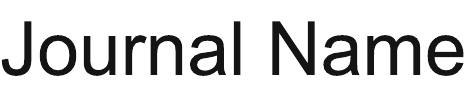}\hfill\raisebox{0pt}[0pt][0pt]{\includegraphics[height=55pt]{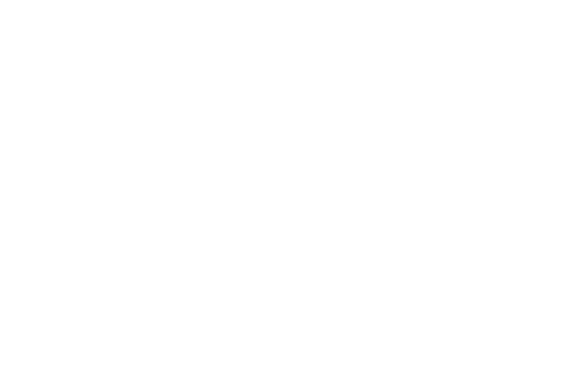}}\\[1ex]
\includegraphics[width=18.5cm]{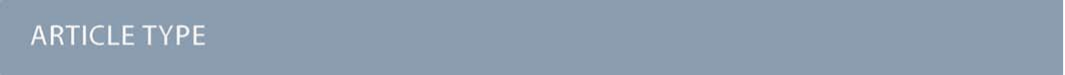}}\par
\vspace{1em}
\sffamily
\begin{tabular}{m{4.5cm} p{13.5cm} }

\includegraphics{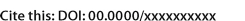} & \noindent\LARGE{\textbf{Tuning field amplitude to minimise heat-loss variability in magnetic hyperthermia}}\\
\vspace{0.3cm} & \vspace{0.3cm} \\

 & \noindent\large{ Necda \c Cam,$^{\ast}$\textit{$^{a,b}$} Iago L\'opez-Vázquez,$^{}$\textit{$^{a,b}$}  \`Oscar Iglesias,\textit{$^{c}$} David Serantes\textit{$^{a,b}$} } \\%Author names go here instead of "Full name", etc.

\includegraphics{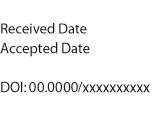} & \noindent\normalsize{ 
In this work, we theoretically investigate how shape-induced anisotropy dispersion and magnetic field amplitude jointly control both the magnitude and heterogeneity of heating in magnetite nanoparticle assemblies under AC magnetic fields. Using real-time Landau–Lifshitz–Gilbert simulations with thermal fluctuations, and a macrospin model that includes both the intrinsic cubic magnetocrystalline anisotropy and a shape-induced uniaxial contribution, we analyze shape-polydisperse systems under clinically and technologically relevant field conditions. We show that for relatively large particles, around 25 to 30 nm, the relative dispersion of local (single-particle) losses exhibits a well-defined minimum at moderate field amplitudes (between 4 to 12 mT), hence identifying an optimal operating regime that minimizes heating heterogeneity while maintaining substantial power dissipation. The position of this critical field depends mainly on particle size and excitation frequency, and only weakly on shape dispersion, offering practical guidelines for improving heating uniformity in realistic MFH systems.
}\\

\end{tabular}

 \end{@twocolumnfalse} \vspace{0.6cm}
]
%%%END OF TITLE, AUTHORS AND ABSTRACT%%%

%%%FONT SETUP - please do not change any commands within this section
\renewcommand*\rmdefault{bch}\normalfont\upshape
\rmfamily
\section*{}
\vspace{-1cm}

%%%FOOTNOTES%%%

\footnotetext{\textit{$^{a}$~Applied Physics Department, Universidade de Santiago de Compostela, Spain}}

\footnotetext{\textit{$^{b}$~~Instituto de Materiais (iMATUS), Universidade de Santiago de Compostela, Spain.}}

\footnotetext{\textit{$^{c}$~Dept. de Física de la Matèria Condensada, Universitat de Barcelona and IN2UB, Barcelona, Spain.}}

%Please use \dag to cite the ESI in the main text of the article.
%If you article does not have ESI please remove the the \dag symbol from the title and the footnotetext below.
%\footnotetext{\dag~Supplementary Information available: [details of any supplementary information available should be included here]. See DOI: 00.0000/00000000.}
%additional addresses can be cited as above using the lower-case letters, c, d, e... If all authors are from the same address, no letter is required

\footnotetext{$^{\ast}$~Corresponding author's email: necda.cam@usc.es}

%%%END OF FOOTNOTES%%%

%%%MAIN TEXT%%%%

\section{Introduction}
%Magnetic fluid hyperthermia (MFH) refers to the phenomena of heat release by magnetic nanoparticles (MNPs) when subjected to an AC magnetic field \cite{perigo2015fundamentals}. While its main focus has been for the treatment of cancer \cite{ivkov2013magnetic,latorre2009applications}, the interest expanded to a vast range of applications, mainly biomedical \cite{rivera2021emerging} but also non biomedical ones, as catalysis \cite{ovejero2021selective} or environmental remediation \cite{baral2024shape}; for a recent review of non-cancer treatment uses of MFH, please refer to ref \cite{gavilan2025}.

Magnetic fluid hyperthermia (MFH) refers to the generation of heat by magnetic nanoparticles (MNPs) subjected to an alternating (AC) magnetic field \cite{perigo2015fundamentals}. Although MFH has been primarily developed for cancer therapy \cite{ivkov2013magnetic,latorre2009applications}, its scope has progressively expanded to a wide range of applications, mainly biomedical \cite{rivera2021emerging}, but also extending to non-biomedical fields such as catalysis \cite{ovejero2021selective} and environmental remediation \cite{baral2024shape}. For a recent review of non-oncological applications of MFH, see Ref.~\cite{gavilan2025}.

%Given that heat dissipation occurs at the single-particle level (mainly at the irreversible magnetisation switch \cite{soetaert2020cancer}), applications based on MFH would mostly benefit from a similar heating behavior of each particle within the system. That is the subject of the present work: to investigate how to minimise the dispersion of energy release (from each particle), considering that experimental systems of MNPs are intrinsically polydisperse \cite{rosensweig2002}, both in size and (more important) in anisotropy. 
Heat dissipation in MFH occurs at the single-particle level, predominantly through irreversible magnetization switching processes \cite{soetaert2020cancer}. As a consequence, the macroscopic heating response of an ensemble reflects an underlying distribution of local (single-particle) energy losses. In practical applications, large particle-to-particle variations in heat generation may lead to %non-uniform temperature distributions, NOT SURE ABOUT THE PREVIOUS AFFIRMATION
local over- or under-heating, and thus reduced efficiency or controllability of the process\cite{munoz2016distinguishing}. From this perspective, MFH-based applications would benefit not only from maximizing the average heating power, but also from minimizing the dispersion of energy dissipation among individual nanoparticles. This consideration motivates the present work, which aims to explore how heating heterogeneity can be reduced in realistic MNP assemblies that are intrinsically polydisperse \cite{rosensweig2002}, both in size and—more critically—in magnetic anisotropy.

The influence of anisotropy dispersion on heating performance is particularly important when compared with size dispersion, because anisotropy plays a dual role in magnetic hyperthermia. Within the commonly accepted picture—where heat dissipation is mainly attributed to irreversible intraparticle magnetization processes \cite{serantes2018anisotropic,di2014magnetic,soukup2015situ}—the particle volume $V$, together with the magnetic anisotropy, determines the height of the energy barrier $E_B$ governing magnetization reversal. %For uniaxial anisotropy, this barrier is given by $E_B = KV$, where $K$ is the uniaxial anisotropy constant. 
Size polydispersity has been shown to affect heating both locally and at the ensemble-averaged level, leading either to a reduction in energy dissipation (by about 30-$50\%$ for a standard deviation $s = 0.20$–$0.25$ \cite{khandhar2011monodispersed}) or, in some cases, to an enhancement of heating performance \cite{jeun2009effects}.

%%% I am here
%The anisotropy, on the other hand, in addition to tuning the energy barrier (and thus the role of temperature), it also defines a field threshold for efficient heat generation: for a randomly dispersed system of MNPs, heating performance \textit{vs.} field amplitude ($H_{max}$) exhibits a sigmoidal shape, evolving from very small values to very large ones at a threshold field that is directly proportional to the so-called \textit{anisotropy field}, $H_K$ \cite{munoz2017towards,ovejero2021selective}. Thus, addressing the role of the anisotropy on heating performance arises as a critical question for MFH applications\cite{vallejo2013effect}.
Magnetic anisotropy, on the other hand, 
also defines a characteristic field threshold for efficient heat generation. For randomly oriented assemblies of MNPs, the heating performance as a function of the field amplitude %($H_{\max}$)
typically exhibits a sigmoidal behavior, evolving from negligible to substantial values above a threshold field proportional to the anisotropy field $H_K$ \cite{munoz2017towards,ovejero2021selective}. As a result, the applied field amplitude emerges as a key experimental control parameter, capable of selectively activating subsets of nanoparticles depending on their anisotropy. Understanding how this field threshold is distributed in polydisperse systems is therefore central to controlling both the magnitude and the homogeneity of heat generation in MFH applications \cite{vallejo2013effect,munoz2017towards}. 

In this work, we investigate the interplay between particle-shape dispersity and field amplitude on local (single-particle) heating performance, focusing on magnetite nanoparticles, the most widely used material in magnetic hyperthermia. There is growing recognition that, for theoretical models to accurately reproduce experimental observations, deviations from ideal spherical particle shapes must be explicitly taken into account \cite{roca2019design,salazar2008cubic,paez2023optimization}. Such deviations, often described as small elongations from sphericity, give rise to a shape-induced uniaxial anisotropy that combines with the intrinsic magnetocrystalline anisotropy to determine the overall magnetic behavior \cite{usov2019heating}. In magnetite, the intrinsic anisotropy is cubic, and neglecting this contribution may lead to an incomplete or misleading description of the heating response, particularly under low-field conditions where the competition between anisotropy terms becomes critical\cite{failde2024}.

Here, we revisit and extend a theoretical prediction previously reported by our group, which showed that the variability of local (single-particle) heating can be reduced by appropriately tuning the field amplitude $H_{\max}$ as a function of the anisotropy dispersion\cite{munoz2017towards}. That earlier study was based on Monte Carlo simulations of systems with only uniaxial anisotropy, and did not account for real-time magnetization dynamics. Building on those results, we now investigate how shape-induced anisotropy polydispersity and field amplitude jointly control the \emph{distribution} of single-particle hysteresis losses under dynamic conditions. Our objective is to identify operating regimes in which heating heterogeneity in shape-polydisperse nanoparticle assemblies is minimized, while maintaining significant overall heat generation. To this end, we employ real-time magnetization dynamics simulations and a comprehensive anisotropy model that combines uniaxial shape anisotropy with the intrinsic cubic magnetocrystalline contribution \cite{failde2024}.

\section{Physical model}

%%% I am here
%We consider magnetite MNPs of diameter $D$ within the macrospin approximation, so that each particle is described by a single magnetic moment $\boldsymbol{\mu}_i = M_s V\,\mathbf{m}_i$, where $M_s$ is the saturation magnetization, $V$ is the particle volume, and $\mathbf{m}_i$ is the unit magnetization vector. Unless otherwise stated, simulations are performed at $T=300$~K.
We consider assemblies of magnetite nanoparticles described within the macrospin approximation, whereby each particle is represented by a single magnetic moment $\boldsymbol{\mu}_i = M_s V\,\mathbf{m}_i$ with $M_s$ the saturation magnetization, $V$ the particle volume, and $\mathbf{m}_i$ a unit vector defining the magnetization direction. The total magnetic energy of each nanoparticle includes three contributions: (i) the intrinsic cubic magnetocrystalline anisotropy; (ii) a uniaxial shape-induced anisotropy, accounting for small deviations from spherical geometry;  and (iii) the Zeeman interaction with the applied time-dependent magnetic field. For magnetite, the cubic anisotropy constant is taken as $K_c = -1.1 \times 10^4$ J m$^{-3}$.

Shape-induced anisotropy is modeled by approximating nanoparticles as prolate ellipsoids with aspect ratio $r = c/a$ ($c$ the long axis, $a=b$ the short axes)\cite{failde2024}. The corresponding uniaxial anisotropy constant is given by
\begin{equation}
K_u = \frac{\mu_0}{2}\left(N_a-N_c\right)M_s^2,
\label{eq:Ku}
\end{equation}
where $N_c$ and $N_a$ are the demagnetizing factors along the long and short axes, respectively. 
Their analytical expressions follow standard results for prolate ellipsoids \cite{bertotti1998hysteresis}.
%They are computed as
%\begin{equation}
%N_c = \frac{1}{r^2-1}\left[\ln\!\left(r+\sqrt{r^2-1}\right)+\sqrt{r^2-1}\right]^{-1},
%\label{eq:Nc}
%\end{equation}
%\begin{equation}
%N_a = \frac{1-N_c}{2}.
%\label{eq:Na}
%\end{equation}

\begin{figure}[H]
    \centering
    \includegraphics[width=1.0\linewidth]{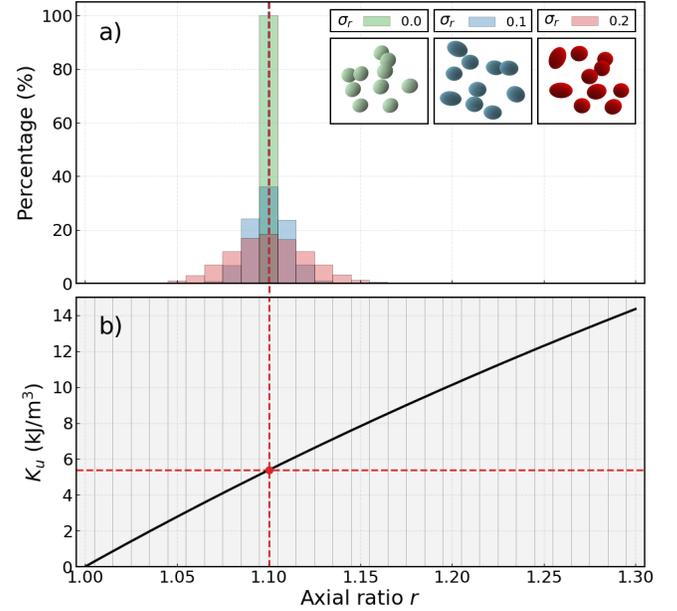}
    \caption{(a) Examples histograms of Gaussian distributions of aspect ratio $r$, centered around the mean $\langle r \rangle=1.1$, for different standard deviations $\sigma_r$, 0.0 (monodisperse) and 0.1 and 0.2. Schematic representations of the corresponding shape-polydisperse ensembles are shown in the upper right corner for illustration purposes.
(b) Dependence of the corresponding uniaxial shape anisotropy constant $K_u$ on $r$. The dashed vertical line highlights the mean axial ratio $\langle r \rangle = 1.1$ of the distributions.}
    \label{Final_Fig_1}
\end{figure}

A key novelty of the present work is that we consider a distribution of shape-induced uniaxial anisotropy constants to be directly correlated with the dispersion in aspect ratios, which we assume to follow a Gaussian distribution. In the present study, we will consider, for all particle sizes, that the system always has a shape dispersity with mean $\langle r \rangle=1.1$ and different standard deviations $\sigma_r$, that will be set to $0.1, 0.2$.
An illustrative scheme of shape distributions and the direct correlation with the associated $K_u$ values is shown in Fig.~\ref{Final_Fig_1}.

%%% I am here
%To account for shape polydispersity, we introduce a distribution of uniaxial shape anisotropy values. Following Mu\~noz-Men\'endez \textit{et al.}~\cite{munoz2017towards}, $K_u$ is sampled from a Gaussian distribution
%\begin{equation}
%f(K;\langle K\rangle;\sigma_K)=\frac{1}{\sqrt{2\pi}\,\sigma_K}\exp\!\left[-\frac{(K-\langle K\rangle)^2}{2\sigma_K^2}\right].
%\label{eq:gaussian}
%\end{equation}
%Here, $\langle K\rangle$ is chosen as $K_u=5371~\mathrm{J/m^3}$ (corresponding to $r=1.1$), and $\sigma_K$ controls the spread. 
%Since $K_u$ is uniquely determined by the aspect ratio $r$, we will also refer to polydispersity in terms of an equivalent shape-dispersion parameter $\sigma_r$.
%To account for shape polydispersity, we introduce a distribution of uniaxial anisotropy values $K_u$ sampled from a Gaussian distribution characterized by a mean value $\langle K_u \rangle$ and standard deviation $\sigma_K$. Since $K_u$ is uniquely determined by the aspect ratio $r$, polydispersity is equivalently expressed in terms of an equivalent shape-dispersion parameter $\sigma_r$.

%For different particle sizes and distribution of aspect ratios, we simulated the usual MFH parameter, the so-called \textit{Specific Loss Power}, SLP, obtained from the area of the $M(H)$ hysteresis loop as
%\begin{equation}
%\mathrm{SLP}=\frac{f}{\rho}\oint M(H)\,dH,
%\label{eq:SLP}
%\end{equation}
%where $f$ is the field frequency and $\rho$ is the material density. The applied field is sinusoidal with amplitude $H_{\max}$.
The heating performance is quantified via the Specific Loss Power (SLP), obtained from the area of the hysteresis loop as
\begin{equation}
\mathrm{SLP}=\frac{f}{\rho}\oint M(H)\,dH,
\label{eq:SLP}
\end{equation}
where $f$ is the field frequency and $\rho$ the material density. The applied magnetic field is sinusoidal, with amplitude $H_{\max}$.

%As a first approximation, for simplicity, in this work we will not consider the role of interparticle interactions. Furthermore, for simplicity we will work under the \textit{frozen ferrofluid} assumption, so that we do not consider rotation of the particles within the fluid, as often considered [refs]. Going beyond these limitations is out of the scope of the present work.
Interparticle interactions and Brownian rotation are not considered in the present study, so that we can isolate the effects of intrinsic anisotropy polydispersity and magnetization dynamics on local heating behavior.

\section{Computational details}
\label{Sec:Compu}
%%% I am here
%We consider a system of 1000 magnetite nanoparticles ($N = 1000$) characterized by diameter $D$ and aspect ratio  $r$. All particles are assumed to lie within the single-domain regime, such that each nanoparticle can be effectively modeled by a single superparamagnetic moment, $\mu_i$. The aspect ratio $r$ of magnetite magnetic nanoparticles has been varied from 1.0 to 1.29, covering the transition from perfectly spherical particles to moderately elongated shapes.
Simulations are performed in the same way as in Refs. \cite{failde2024,ovejero2021selective}, in which the magnetic evolution of a system of magnetite nanoparticles under a time-varying magnetic field is simulated with the OOMMF micromagnetic package, which numerically integrates the Landau–Lifshitz–Gilbert (LLG) equation \cite{donahue1999oommf}, with a random field to account for thermal effects\cite{lemcke2004thetaevolve}. Each nanoparticle is modeled as a single discretized cubic cell, whose volume is taken equal to the particle volume under the macrospin approximation. All the simulations are carried out for a system of $N=1000$ particles at $T = 300$ K, using a Gilbert damping parameter $\alpha = 0.1$. %\textcolor{red}{I DON'T THINK THIS SENTENCE IS NEEDED; IF SO, SAYING AS IN OUR PREVIOUS WORK, REF. \cite{failde2024}The equivalent spherical diameter is thus given by $D = 1.24\,L$, where $L$ is the cell size.}
To obtain the heating performance, different hysteresis loops are simulated until convergence is reached, and then the hysteresis area is evaluated to obtain the SLP through Eq.~\ref{eq:SLP}. 

The present results correspond always to size-monodisperse cases, to focus on the role of shape (anisotropy) dispersity. Yet, to assess size effects, simulations have been performed for selected particle diameters $D = $15, 20, 25, and 30 nm. The aspect ratio $r$ is varied between $1.0$ and $1.29$, spanning the transition from spherical to moderately elongated particles. It is important to keep in mind that in each simulation the particle volume is the same for every particles, independently of its elongation; the diameter $D$ is reported solely as a reference, for the ideally spherical particle equivalent. The orientations of both uniaxial and cubic anisotropy axes are randomly assigned.
% ?? %The Gilbert damping parameter was chosen as $\alpha = 0.1$. Each particle was modeled as a discretized cubic cell, with the cell volume $L^3$ assumed to be equal to the nanoparticle volume under the macrospin approximation. Consequently, for a given cell size $L$, the diameter of a sphere with the same volume is $D = 1.24L$. 
%%\textcolor{red}{THIS WAS MENTIONED IN SEC.2
%%To account for shape polydispersity, we introduce a distribution of uniaxial anisotropy values $K_u$ sampled from a Gaussian distribution characterized by a mean value $\langle K_u \rangle$ and standard deviation $\sigma_K$. Since $K_u$ is uniquely determined by the aspect ratio $r$, polydispersity is equivalently expressed in terms of an equivalent shape-dispersion parameter $\sigma_r$, that will be set to $0.1, 0.2$.}
For each set of parameters, 30 independent realizations are simulated to ensure statistical robustness.
%For the calculation of the SLP in Eq.\ref{eq:SLP}, the mass density of magnetite was set to $\rho = 5170~\mathrm{kg\,m^{-3}}$. Using the analytical formula given in Eq.\ref{eq:gaussian}, 30 independent systems were considered, where the aspect ratio $r$ has varied from 1.0 to 1.29 in steps of 0.01 for a uniaxial shape anisotropy along a single axis. The cubic anisotropy constant $K_C$ was kept fixed, with a value of $K_C = -1.1 \times 10^4~\mathrm{J/m^3}$ for magnetite. The orientations of the axes were randomly assigned for both the uniaxial and cubic anisotropies. For magnetite, a saturation magnetization of $M_s = 480 \times 10^3~\mathrm{A/m}$  has been selected.
%A time-dependent magnetic field, $B_x(t) = B \sin(\omega t)$, was applied along the $x$-direction. Here, $t$ denotes time and $\omega = 2\pi f$ is the angular frequency, where $f$ is the field frequency expressed in units of kHz. Frequencies of $f = 1000~\mathrm{kHz}$ and $f = 100~\mathrm{kHz}$ were considered. In general, a nanoparticle diameter of $D = 25$~nm has been assumed; however, to investigate the effect of particle size on local hysteresis losses in Sec.\ref{heat production}, simulations were also performed for $D = 15$, 20, and 30~nm. The standard deviation parameter $\sigma_r$ was set to $\sigma_r = 0$ for the monodisperse system, while values of $\sigma_r = 0.1$ and $0.2$ were used for the polydisperse systems, as described in Eq.\ref{eq:gaussian}.
A time-dependent magnetic field $H(t) = H_{\max}\sin(2\pi f t)$ is applied along a fixed spatial direction, considering two frequencies, $f = 100$ kHz and $f = 1000$ kHz. The material parameters for magnetite are $M_s = 4.8 \times 10^5$ A m$^{-1}$ and $\rho = 5170$ kg m$^{-3}$.

%%% I am here: Secs. 1, 2, 3 Revised by Òscar Jan 14th
\section{Results and discussion}

\begin{figure}[H]
    \centering
    \includegraphics[width=1.0\linewidth]{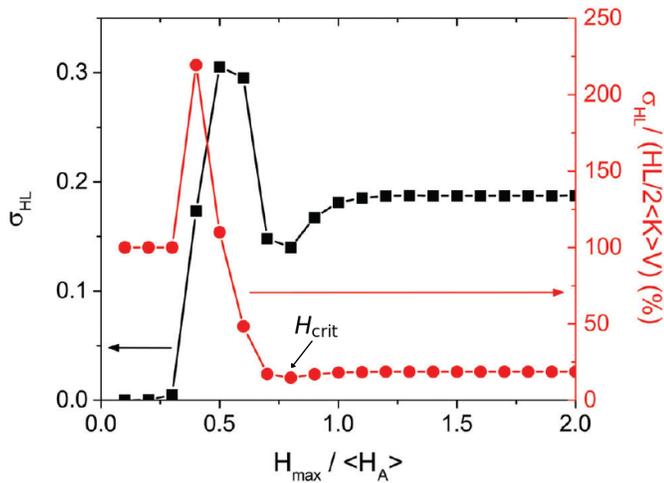}
    \caption{Left axis: Standard deviation of the normalized local hysteresis
losses $\sigma_{HL}$, as a function of the normalised applied field amplitude, $H_{max}/\langle{H_A\rangle}$, for a distribution in anisotropy constants $K_u$ of of $\sigma = 0.2$. Right axis: Percentage of this standard deviation in relation to the global normalized hysteresis losses. The $H_{crit}$ value indicates the critical field for minimising dispersion in heat losses. Adapted from Ref. ~\cite{munoz2017towards} with permission from the Royal Society of Chemistry.}
    \label{Final_Fig_2}
\end{figure}

Munoz-Menendez and colleagues ~\cite{munoz2017towards} have demonstrated that, for blocked-like particles, the variation in local (single-particle) hysteresis losses can be controlled by adjusting the maximum magnetic field intensity $H_{\max}$ in accordance with the anisotropy polydispersity distribution parameter, $\sigma_r$, using Monte Carlo simulations; see Fig.~\ref{Final_Fig_2}, reproduced from Ref. \cite{munoz2017towards}.

The standard deviation of the local losses, $\sigma_{HL}$, \textit{vs.} $H_{\max}$ (left axis in Fig. \ref{Final_Fig_2}), might apparently suggest that the smallest fields are the more adequate to diminish the dispersion in local heating. However, since at very small fields the heating is also very small, it is more appropriate to analyse the relative importance $\sigma_{HL}$ against the correspondent global losses, as shown by the right axis in Fig. \ref{Final_Fig_2}. In this case, it is clear that at low fields, the distribution of local hysteresis losses is quite significant compared to the relatively low global losses. In contrast, at higher fields, the dispersion of local losses diminishes significantly, showing a minimum at $\sim{H_{\max}=0.8 H_K}$. The key objective of the present work is to investigate the possible existence of this optimising field amplitude, under the more realistic conditions of the present model. %To this end, first we will consider the role of shape dispersion on global losses at the global (entire system) level, in Subsection \ref{subsec:global}, both for shape-monodisperse ($\sigma_r=0.0$; \ref{subsubsec:shape_mono}) and polydisperse condidions ($\sigma_r>1; \ref{subsubsec:shape_poly}$). Then, in subsection \ref{subsec:local}, we analyse the dispersion in local heating performance as a function of $H_{max}$, $\sigma_r$, and $D$.

%\textcolor{red}{This has already been said at the end of Sec. 3. Can be deleted In this study, we provide a comprehensive analysis of the SLP behavior in monodisperse and polydisperse nanoparticles by incorporating both uniaxial shape anisotropy and a constant cubic magnetocrystalline anisotropy, using real-time dynamical simulation.
%}

\subsection{Global (entire system) heating: role of aspect ratio}\label{subsec:global}

%As introduced previously, we aim to analyze how the nanoparticles aspect ratio influences the global heating performance of the system. 

We first consider shape-monodisperse assemblies ($\sigma_r=0.0$; part \ref{subsubsec:shape_mono}), in which all nanoparticles have the same axial ratio $r$, to establish a direct relationship between shape anisotropy and heating performance. We then extend the analysis to shape-polydisperse systems ($\sigma_r>1$; part \ref{subsubsec:shape_poly}), where a distribution of aspect ratios is introduced to account for realistic nanoparticle ensembles. In this second stage we shall always restrict to the case $\langle r \rangle=1.1$ as a reference sample of what is usually considered (quasi)homogeneous experimentally, as ideal perfectly spherical (or cubic) particles cannot be synthesised yet\cite{lopez2025role}. %This second step enables us to assess how shape-induced disorder affects both the global SLP and the dispersion of local heat production within the system.

\subsubsection{Shape-monodisperse system}\label{subsubsec:shape_mono}

\begin{figure}[H]
    \centering
    \includegraphics[width=1.0\linewidth]{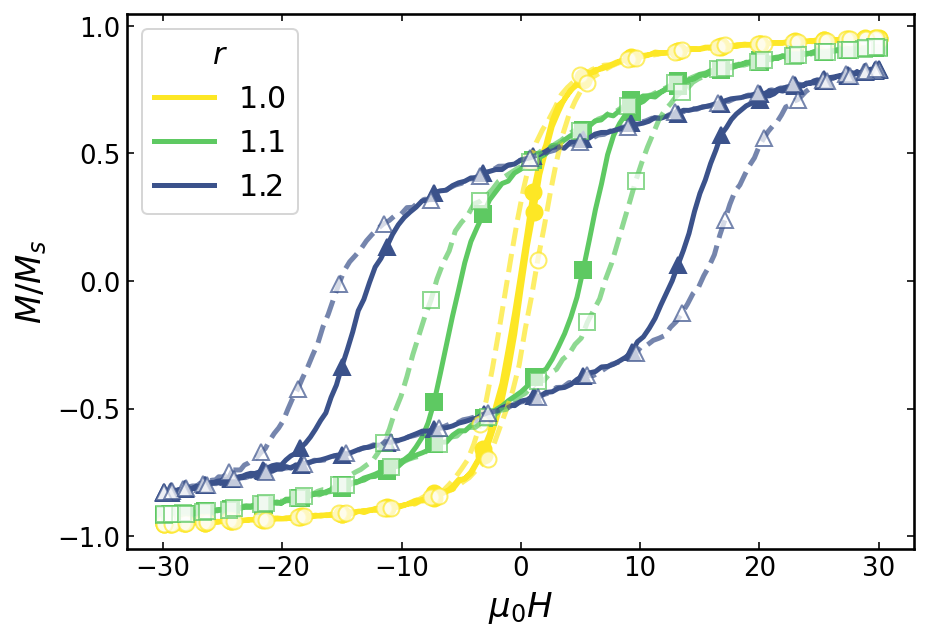}
    \caption{Hysteresis loops for $D=25$ nm particles with different aspect ratios $r$. Solid lines correspond to $f=100$~kHz and dashed lines to $f=1000$~kHz, in both cases for $\mu_0 H_{\max}=30$ mT.}

    \label{Final_Fig_3}
\end{figure}

The influence of the aspect ratio on the magnetic response is first assessed through the hysteresis loops obtained under alternating magnetic fields. Representative loops for monodisperse systems with different aspect ratios $r$ are shown in Fig.~\ref{Final_Fig_3} for a 30\,mT sinusoidal field at $f=100$ and 1000\,kHz. %\textcolor{red}{THE DETAILS ABOUT HEATING ARE DISCUSSED AGAIN IN MORE DETAIL REGARDING FIG. 4 For spherical particles ($r=1.0$), the response at 100\,kHz is nearly reversible, leading to negligible hysteresis loss, whereas at 1000\,kHz the magnetization cannot fully follow the field and the loop opens. Increasing $r$ increases the uniaxial shape anisotropy, which raises the reversal energy barrier and enhances dynamic losses; consequently, the loop area $H_L$ increases. Overall, hysteresis becomes more pronounced at higher frequencies and for larger $r$ values.}

% Hysteresis losses are closely related to the SLP behavior.

% 

\begin{figure}[H]
    \centering    
    \includegraphics[width=1.0\linewidth]{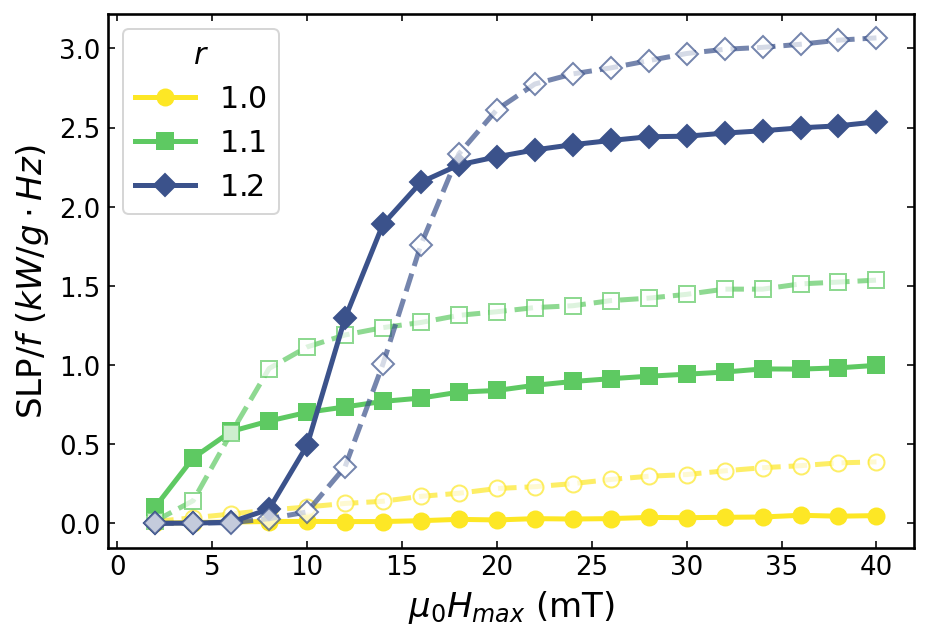}
    \caption{SLP/f \textit{vs.} $\mu_0 H_{\max}$ for $D=25$ nm particles, for systems with different aspect ratios $r$; for both $f=100$~kHz (solid lines) and $f=1000$~kHz (dashed lines).}
    \label{Final_Fig_4}
\end{figure}

Two main features are observed in Fig.~\ref{Final_Fig_3}. First, increasing $r$ changes the general shape of the curves, from very narrow and quasi-saturated at $r=1.0$, widening and deviating from saturation with growing $r$. Second, increasing $f$ does not significantly change the shape of each $r$ case, only slightly opening the loop. 
These changes in the magnetic hysteresis directly translate into variations of the heating efficiency. The dependence of the SLP on the applied field amplitude $H_{max}$ for different aspect ratios and frequencies is presented in Fig.~\ref{Final_Fig_4}. For clarity, the results are reported as $\mathrm{SLP}/f$, which allows a direct comparison between the $f=100$ and $1000$~kHz cases, as estimates of the heating performance per cycle. 

It is can be seen that variations in $r$ lead to markedly different heating performances, with clearly distinct behaviour in the low and high field regimes. 
At high fields, a general trend of increasing $SLP/f$ values with increasing $r$ is observed.
In contrast, the low-field behaviouris more intricate: significant heating occurs only for $r=1.1$, while it remains negligible for both $r=1.0$ and $r=1.2$.
Overall, increasing the frequency results in non-negligible losses for the ideal case $r=1.0$ and accentuates the sigmoidal character of the curves for $r>1$. A detailed explanation of these features can be found in Refs.~\cite{ovejero2021selective,failde2024}, in terms of the relative contributions of the intrinsic cubic magnetocrystalline anisotropy and the shape-uniaxial one to the energy barriers.

Briefly, at $r=1.1$ the uniaxial shape-anisotropy contribution becomes comparable to the cubic one, increasing the effective barrier and hindering the ability of the magnetization to follow the alternating field. Consequently, hysteresis losses increase and measurable $\mathrm{SLP}/f$ values appear already at low field amplitudes (about 4~mT at 100~kHz). The crossover to uniaxial shape-anisotropy dominance occurs at $r\simeq1.22$~\cite{failde2024}; therefore, at $r=1.22$ the effective barrier is substantially larger (12 times higher than the cubic barrier), and appreciable losses are observed only above higher amplitudes (around 10~mT at 100~kHz). Beyond $\sim$15~mT, $\mathrm{SLP}/f$ tends to increase more moderately. Increasing the frequency further enhances dynamic losses, yielding the largest $\mathrm{SLP}/f$ for $r=1.2$ at 1000~kHz.

\subsubsection{Shape-polydisperse system}
\label{subsubsec:shape_poly}

We now consider polydispersity in the uniaxial anisotropy values to account for shape-induced variability, as presented in Fig.\ref{Final_Fig_1}. Since we assume non-interacting particles, to construct polydisperse ensembles we can simply weight the relative contribution of each fraction of particles with a given $r$ according to the desired $\sigma_r$ distribution. Thus, in the simulations we followed the same procedure described in Sec.~\ref{subsubsec:shape_mono}, for different values of $r$. Specifically, we carried out $30$ independent simulations for each $\sigma_r$ value, varying $r$ from $1.0$ to $1.29$ in steps of $0.01$. For each $r$, hysteresis loops and the corresponding SLP values were computed by sweeping $\mu_0H_{max}$ from $1$ to $10$~mT in $1$~mT increments, and from $10$ to $40$~mT in $2$~mT increments. This procedure was repeated for the different nanoparticle diameters $D$. It is important to recall that for all cases $\langle r \rangle=1.1$, as stated earlier.

\begin{figure}[H]
        \centering
        \includegraphics[width=1.0\linewidth]{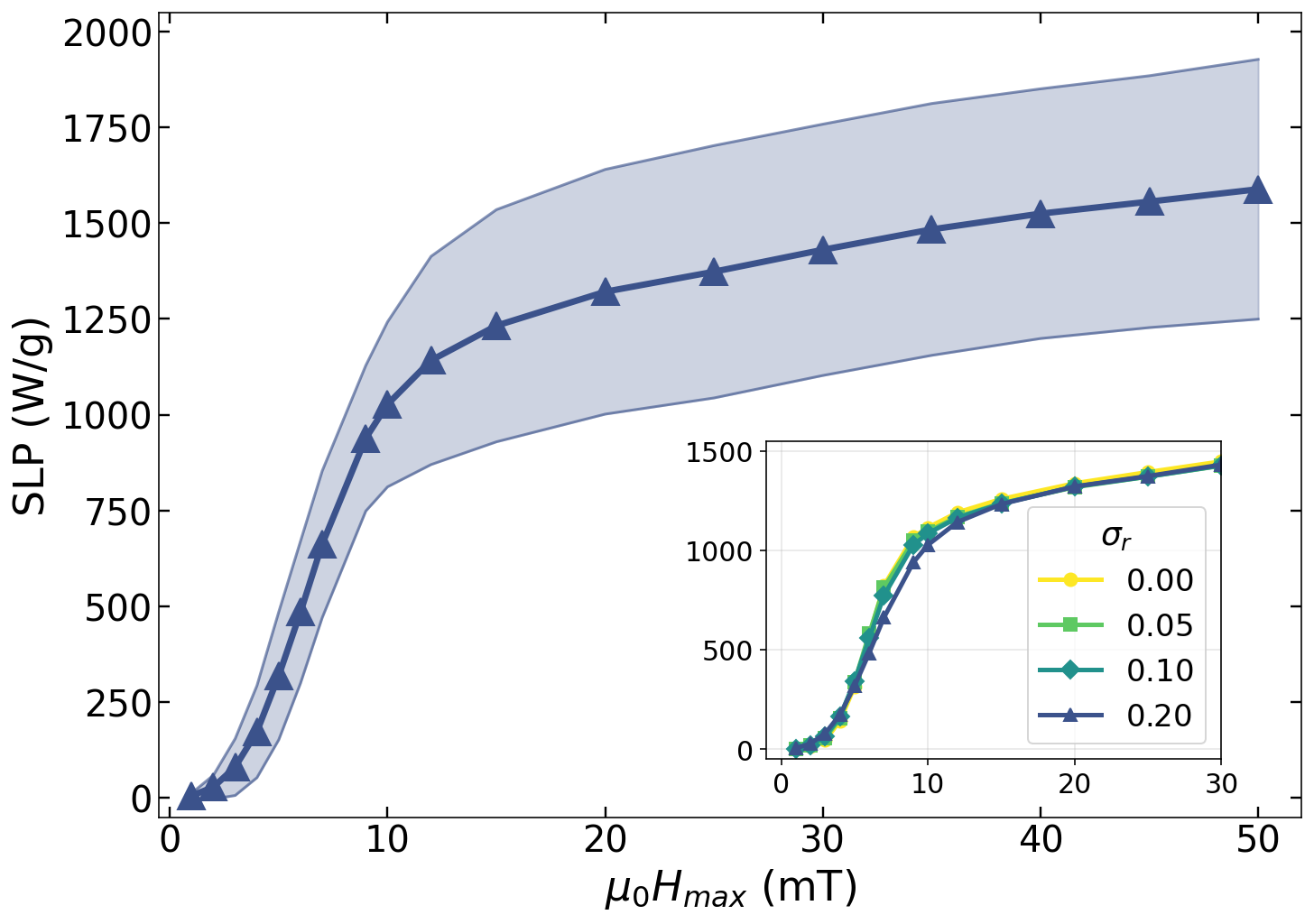}
        \caption{SLP \textit{vs.} $\mu_0 H_{\max}$ for shape-polydisperse MNPs with $\sigma_r = 0.2$, $f = 1000$~kHz, and $D = 25$~nm. The symbols represent the ensemble-averaged SLP, while the shaded region stands for the standard deviation of SLP, $\sigma_{SLP}$, arising from the distribution of particle shapes. The inset shows the SLP value as a function of $\mu_0 H_{\max}$ for other $\sigma_r$ cases.
}
        \label{Final_Fig_5}
   \end{figure}

For a given degree of shape polydispersity, the effective SLP of the ensemble was obtained by weighting the individual SLP contributions corresponding to each axial ratio according to the Gaussian distribution defined by the chosen value of $\sigma_r$. In this way, the resulting SLP captures the combined response of nanoparticles with different shapes within the ensemble. To quantify the dispersion, we define the standard deviation of the local SLP values as
\begin{equation} \label{eq:sigma_slp}
\sigma_{SLP} = \sqrt{\frac{1}{N} \sum_{j=1}^{P} N_j \left( SLP_j - \langle SLP \rangle \right)^2},
\end{equation}
where $P$ denotes the number of particle categories, $N$ is the total number of nanoparticles in the ensemble, and $SLP_j$ corresponds to the SLP associated with the $j$-th category. As an illustrative example, Fig.~\ref{Final_Fig_5} shows the SLP obtained for the case $\sigma_r = 0.2$ as a function of $\mu_0 H_{\max}$. The shaded region represents the standard deviation of the SLP values.

Two main features are observed in Fig. \ref{Final_Fig_5}. First, the overall shape of the SLP \textit{vs.} $\mu_0 H_{\max}$ curve closely resembles that of the shape-monodisperse systems with $r>1.0$. Second, the width of the $\sigma_{SLP}$ value increases significantly as the SLP tends to saturation. Both features originate from the progressive contribution of particles with 
larger shape anisotropy as $H_{\max}$ increases.

%On the one hand, the overall shape of the SLP \textit{vs.} $B_{max}$ being similar to the shape-monodisperse $r=1.1$ (and $r=1.2$) cases indicates a dominance of the more elongated particles (i.e. higher uniaxial shape anisotropy) on the global SLP value of the entire system. This interpretation is supported by the rough overlap of the $\sigma_r$ cases (inset in Fig. \ref{fig5}); please note that in fact, the $\sigma_r=0$ case in Fig. \ref{fig5} is in fact the $r=1.1$ case in Fig. \ref{fig3} (where $\sigma_r=0$).
The first feature reflects the dominant role of the more elongated particles
%—those with larger uniaxial shape anisotropy—
in determining the ensemble response. Accordingly, the polydisperse curves closely follow the monodisperse $r=1.1$ and $r=1.2$ cases, and the different $\sigma_r$ curves nearly overlap (inset of Fig.~\ref{Final_Fig_5}); note that the $\sigma_r=0$ curve in Fig.~\ref{Final_Fig_5} corresponds to the monodisperse $r=1.1$ case shown in Fig.~\ref{Final_Fig_4}.

The pronounced increase in $\sigma_{SLP}$ near the saturation regime arises from the progressive activation of particles with higher uniaxial shape anisotropy as $H_{\max}$ increases. As discussed in Ref.~\cite{munoz2017towards}, particles with larger $K_u$ dissipate more energy but require larger fields to reach major‑loop conditions. Increasing $H_{\max}$ therefore enables these particles to contribute, simultaneously enhancing the global SLP and broadening the local losses.

A simplified, more intuitive explanation of this threshold at the saturation part of the curve follows from the crossover between the $r=1.1$ and $r=1.2$ curves in Fig. \ref{Final_Fig_4}, which occurs at $\mu_0 H_{\max}\sim{14}$ mT: before this field, the highest contribution to the SLP is from the $r=1.1$ case, and the deviation in losses is more moderate (both more and less elongated particles give very small SLP), whereas above it the $r=1.2$ contributes more strongly, raising the average SLP while simultaneously broadening the distribution in local SLP values. 

The results displayed in Fig. \ref{Final_Fig_5} (similar trends are obtained for other $\sigma_r$ and $D$, not shown) indicate that global SLP and its dispersion are not directly correlated.
While the average SLP increases monotonically with $H_{\max}$,
$\sigma_{SLP}$ also grows, so that the latter alone does not provide a reliable measure of the relevance of local heating effects. Identifying a more appropriate descriptor is the subject of the following subsection.
%since the global (average) SLP grows with $B_{max}$ but also also $\sigma_{SLP}$ does, the latter alone does not provide a faithful measure of the relevance of local heat production effects. Finding a more appropriate estimate is the subject of the following subsection.

%For completeness, the inset in Fig.~\ref{fig5} shows the evolution of the SLP as a function of the applied magnetic field for different values of the shape polydispersity parameter $\sigma_r$. All cases exhibit a similar qualitative trend, indicating that the overall field dependence of the SLP is largely preserved across different degrees of shape polydispersity.

\subsection{Dispersion in local (single-particle) heat production} \label{subsec:local}

To analyse local heating effects in a physically meaningful way, we focus on the dispersion normalized by the ensemble-averaged SLP, $\sigma_{\mathrm{SLP}}/\langle \mathrm{SLP} \rangle$. In contrast to the $\sigma_{\mathrm{SLP}}$ value alone, this dimensionless quantity directly quantifies the \emph{relative} heterogeneity of single-particle heat production within the ensemble, and therefore provides an appropriate descriptor for assessing the relevance of local losses. This approach follows the same rationale adopted by Muñoz-Menéndez \textit{et al.}~\cite{munoz2017towards}, who analysed the dispersion of local hysteresis losses relative to the corresponding global losses. The difference is that in our present work we can consider real-time dynamics through the LLG equation, which were not accessible through the Metropolis Monte Carlo method of Ref. \cite{munoz2017towards}.

\subsubsection{Comparison with previous work}\label{subsubsec-comparison}

\begin{figure}[H]
    \centering
    \includegraphics[width=1.0\linewidth]{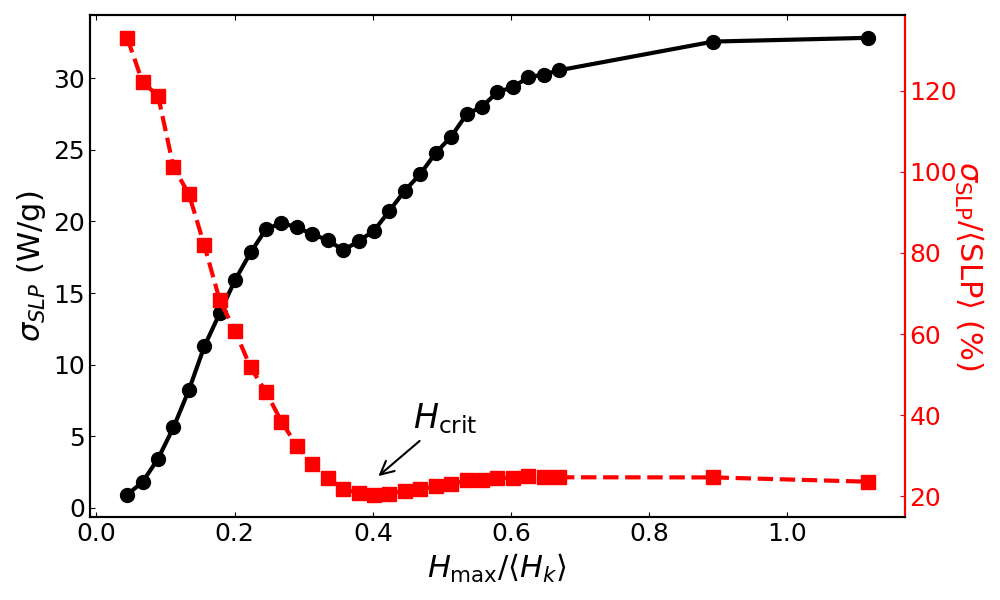}
    \caption{Absolute values of heat-losses dispersion, $\sigma_{SLP}$ (black circles, left axis) and normalised correspondent, $\sigma_{\mathrm{SLP}}/\langle \mathrm{SLP} \rangle$ (red squares, right axis), as a function of $H_{\max}/\langle{H_K}\rangle$, for $D = 25$~nm, $f = 1000$~kHz, and $\sigma_r = 0.2$. The $H_{crit}$ value indicates the critical field for minimising SLP dispersion.}

    \label{Final_Fig_6}
\end{figure}

We first compare the results of our dynamic model against the main result reported in Fig.\ref{Final_Fig_2}.  Fig.~\ref{Final_Fig_6} shows the analogue of the original plot for the same dispersion parameter, $\sigma_r = 0.2$, now formulated in terms of the SLP instead of the hysteresis losses $HL$. Note that in Ref.~\cite{munoz2017towards} no particle size was specified, implicitly assuming that the model applies to blocked-like particles, as reported elsewhere for the same technique \cite{serantes2010influence}. Blocked-like behaviour can be safely expected for magnetite particles with $D=25$ nm at $f=1000$ kHz, as the case shown in Fig.~\ref{Final_Fig_5}. 
To enable a more direct comparison with Fig.~\ref{Final_Fig_2}, $H_{max}$ has also been normalised by the average anisotropy field of the particles, $\langle{H_K}\rangle$; for $\sigma_r=0.2$ and $\langle r \rangle$ = 1.1, $\mu_0\langle{H_K}\rangle=22.38$ mT.

Fig.~\ref{Final_Fig_6} shows an overall increase of $\sigma_{SLP}$ \textit{vs.} $H_{\max}$, as expected (note that this data corresponds to the shaded area in Fig.~\ref{Final_Fig_5}). Remarkably, a secondary minimum is observed at $H_{max}\sim{0.4\langle{H_K}\rangle}$, which was not evident from Fig.~ \ref{Final_Fig_5}. Importantly, such feature is also present in the $\sigma_{HL}$ data of Fig.~\ref{Final_Fig_2}, which also follows an overall increase with the presence of a secondary minimum at intermediate fields. Notable differences are that in the original case an initial constant (nearly-zero) range was followed by an absolute maximum, whereas in the present work no constant range is observed, and the maximum is secondary. In addition, the (normalised) field values at which the minimum appears are significantly smaller for the $\sigma_{SLP}$ case.

In contrast, the normalized quantity $\sigma_{SLP}/\langle{SLP}\rangle$ exhibits a markedly different behaviour. It is largest at low fields (indicating strong heterogeneity in local heating) and decreases rapidly until reaching a minimum at a critical field $H_{\mathrm{crit}}\sim{0.4\langle{H_K}\rangle}$, where heat production becomes more homogeneous. A subsequent weak increase is observed for higher $H_{max}$, followed by a plateau for $H_{max}\gtrsim{0.6\langle{H_K}\rangle}$. This confirms that, while $\sigma_{\mathrm{SLP}}$ alone is not a suitable descriptor (as anticipated from Fig. ~\ref{Final_Fig_5}), the normalized quantity $\sigma_{\mathrm{SLP}}/\langle \mathrm{SLP} \rangle$ captures the existence of an optimal field regime, in direct analogy with Ref.~\cite{munoz2017towards}. 
A key difference, however, is that in the original case a similar trend emerged only after an initial constant regime, which is entirely absent in the present data.
%In comparison with the $\sigma_{HL}$ data of Fig. \ref{Final_Fig_2}, though, there is a significant difference: in the original case a very similar trend was also observed, but after an initial constant range which is totally absent in the current data.

The main differences between our present results and those reported in Ref.~\cite{munoz2017towards} can be attributed to both the different physical model and the computational approach. 
In particular, in the low-field regime the key difference arises from the inclusion of cubic anisotropy in the present work, which enables dissipation at much smaller fields\cite{failde2024}. Thus, with increasing $H_{max}$, there is a progressive contribution of the total losses from particles with higher uniaxial-shape anisotropy, which can undergo magnetisation reversal via the \textit{effective} reduction of the uniaxial energy barrier by the cubic contribution. In contrast, the uniaxial-only anisotropy particles considered in Ref.~\cite{munoz2017towards} required substantially larger fields to dissipate energy, resulting in negligible and nearly constant losses at small $H_{max}$ values. 
In addition, part of the difference arises from the computational methodology itself. While Ref.~\cite{munoz2017towards} employed a Metropolis Monte Carlo scheme based on energy minimization and thus probing quasi-static configurations, our approach relies on the numerical integration of the Landau–Lifshitz–Gilbert equation and explicitly accounts for real-time magnetization dynamics. This allows us to capture frequency-dependent and transient effects in the losses, which become particularly relevant in the low-field regime and further contribute to the absence of an initial constant-loss region in the present results.
% VOU AQUI, tengo que acabar de pulirlo
%Furthermore, this behaviour was reinforced by the use of the energy-minimisation procedure of the Metropolis Monte Carlo technique.\textcolor{red}{Òscar, please check the previous sentences... I wanted to add something regarding the technique too, not only regarding the model...}

\subsubsection{Determination of optimum field conditions}\label{subsubsec-critical fields}

A key outcome of the previous analysis is that the more comprehensive model employed in the present work reproduces the occurrence of the critical field $H_{crit}$ reported in Ref.~\cite{munoz2017towards}. This provides an independent validation of those findings using a more robust computational procedure and a more sophisticated physical description.  In the following, we extend that analysis to consider the role of particle size ($D$), frequency ($f$), and shape polydispersity ($\sigma_r$).

\begin{figure}[H]
    \centering
    \includegraphics[width=1.0\linewidth]{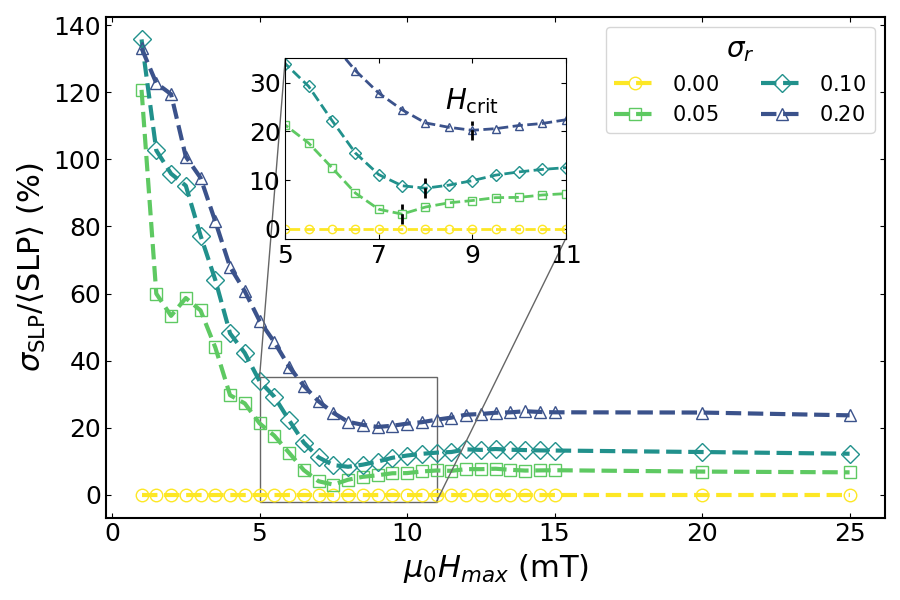}
    \caption{
    $\sigma_{\mathrm{SLP}} / \langle \mathrm{SLP} \rangle$ \textit{vs.} $\mu_0 H_{\max}$ for $f = 1000~\mathrm{kHz}$ and $D = 25~\mathrm{nm}$. The inset highlights the corresponding critical fields $H_{crit}$.
    }
    \label{Final_Fig_7}
\end{figure}

We begin with a representative case, $D=25$~nm and $f=1000$~kHz, and analyse the behaviour of $\sigma_{\mathrm{SLP}}/\langle \mathrm{SLP} \rangle$ as a function of the field amplitude $\mu_0 H_{\max}$ for several values of $\sigma_r$ (Fig.~\ref{Final_Fig_7}).

Figure~\ref{Final_Fig_7} shows a qualitatively similar behaviour for all shape-polydisperse samples (except for the idealized and unrealistic monodisperse limit $\sigma_r=0$). In all cases, $\sigma_{\mathrm{SLP}}/\langle \mathrm{SLP}\rangle$ decreases from a pronounced maximum at low field amplitudes to an absolute minimum around $7$--$9$~mT, followed by a slight increase and a broad plateau at higher fields. The main effect of increasing $\sigma_r$ is a progressive upward shift of the curves, i.e., larger values of $\sigma_{\mathrm{SLP}}/\langle \mathrm{SLP}\rangle$ across the full field range, together with a displacement of the minimum towards larger $\mu_0 H_{\max}$. In particular, this implies a shift of $H_{crit}$ towards larger values as shape dispersion increases.

\begin{figure}[H]
    \centering
    \includegraphics[width=1.0\linewidth]{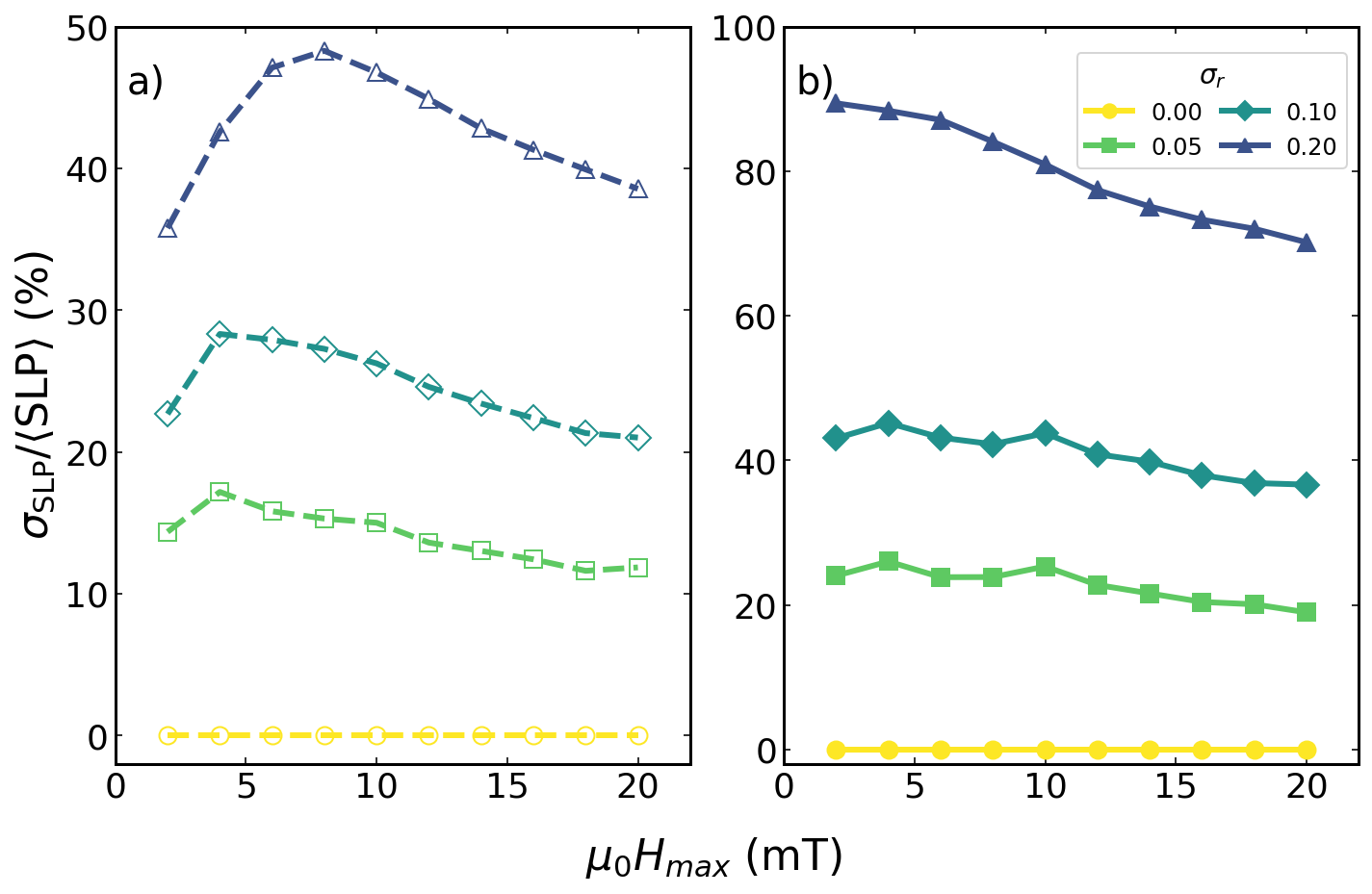}
    \caption{Variation of $\sigma_{\mathrm{SLP}} / \langle \mathrm{SLP} \rangle$ as a function of $\sigma_r$, for $D=20$ nm , respectively. Empty symbols (dashed lines) correspond to $f=1000$~kHz a), and filled symbols (solid lines) to $f=100$~kHz b).}
    \label{Final_Fig_10}
\end{figure}

Although not shown here, a similar trend in the $\sigma_{\mathrm{SLP}} / \langle \mathrm{SLP} \rangle$ \textit{vs.} $\sigma_r$ is observed for the same particle size at $f=100$~kHz, and the overall behaviour remains comparable for $D=30$~nm at both frequencies. However, for smaller particles ($D = 15$ and $20$~nm), the $\sigma_{\mathrm{SLP}} / \langle \mathrm{SLP} \rangle$ \textit{vs.} $H_{max}$ curves change significantly. This is illustrated in Fig.~\ref{Final_Fig_10} for $D=20$~nm. 

In Fig. \ref{Final_Fig_10} it is observed an -apparently- very different behaviour with respect to Fig. \ref{Final_Fig_7}, as no minimum can be identified. For $f=1000$~kHz, the curves display a well-defined maximum within the same field range where the $D=25$~nm case exhibited a minimum. For $f=100$~kHz, the dispersion shows only a weak decrease with increasing $H_{\max}$. In both cases, $\sigma_{\mathrm{SLP}}/\langle \mathrm{SLP}\rangle$ remains substantially larger (by roughly a factor of $2$--$3$) than the minimum values obtained for the larger nanoparticles. The results for the $D = 15$ case (not shown) are very similar to the $f=100$ case of Fig. \ref{Final_Fig_10} for both frequencies, with a smooth decrease with increasing $H_{max}$ for both frequencies. Likewise, for ideal monodisperse ($\sigma_r \equiv 0$) and even nearly monodisperse systems ($\sigma_r = 0.01$), the dispersion does not exhibit a clear minimum, preventing the definition of a critical magnetic field.

\begin{figure}[H]
    \centering
    \includegraphics[width=1.0\linewidth]{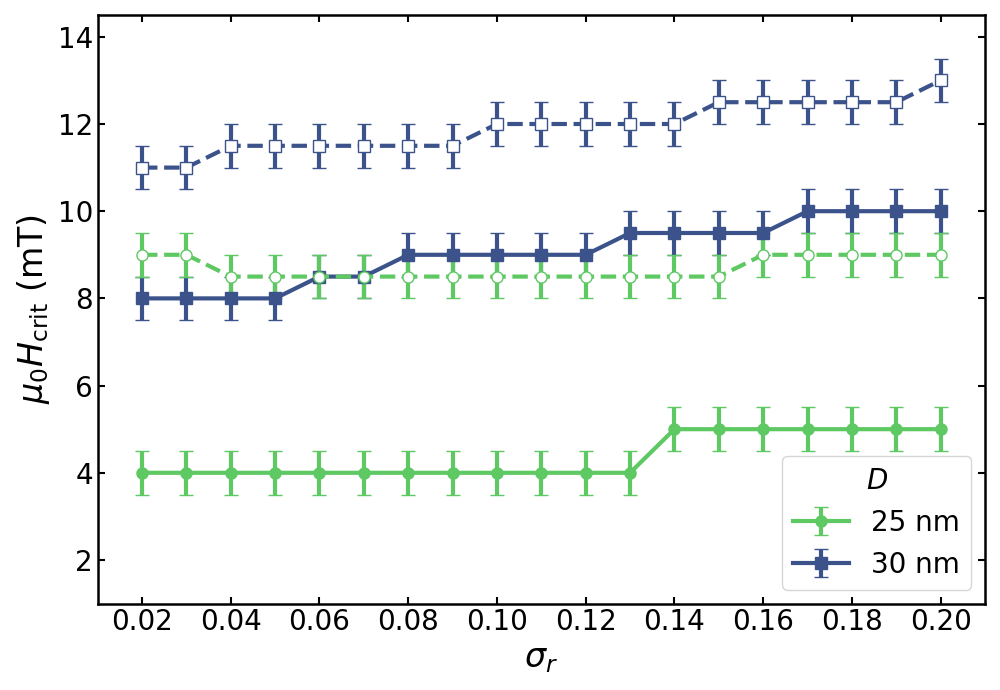}
    \caption{Critical magnetic field $H_{\mathrm{crit}}$ as a function of $\sigma_r$, for $D=30$ nm (blue squares) and $D=25$ nm (orange circles), respectively. Empty symbols (dashed lines) correspond to $f=1000$~kHz, and filled symbols (solid lines) to $f=100$~kHz.}
    \label{Final_Fig_8}
\end{figure}

One might hypothesize that what is observed for the $D=20$ nm case in Fig. \ref{Final_Fig_10} is in fact not that different from the $25$ nm case of Fig. \ref{Final_Fig_7}, and that the curves would eventually show the same minima, just at much larger fields. However, the comparison between the $D=25$ and $D=30$ nm cases does not support that hypothesis, as will be shown later in Fig. \ref{Final_Fig_8}. Furthermore, even if that would be the case, what is clear is that size polydispersity is showing a key role regarding local-heating dispersity, as discussed elsewhere\cite{munoz2015role}. 

Nevertheless, since studying the role of polydispersity lies outside the scope of the present work, in the following we will focus our analysis on the cases where $H_{\mathrm{crit}}$ is well defined, namely $D=25$ and $30$~nm for sufficiently broad shape dispersions. The corresponding values of $H_{\mathrm{crit}}$ as a function of $\sigma_r$ for $f=100$ and $1000$~kHz are summarized in Fig.~\ref{Final_Fig_8}.
At $f = 1000$~kHz, $H_{\mathrm{crit}}$ lies in the range $11$–$13$~mT for $D = 30$~nm, while for $D = 25$~nm it is shifted to lower values, around $8$–$9$~mT, indicating a clear dependence on particle volume. Reducing the particle diameter progressively shifts the minimum towards lower field amplitudes and ultimately suppresses its appearance. Moreover, decreasing the excitation frequency further lowers the critical field: at $f = 100$~kHz, $H_{\mathrm{crit}}$ is found around $4$–$6$~mT for $D = 25$~nm and $8$–$10$~mT for $D = 30$~nm. Overall, these results show that the emergence of $H_{\mathrm{crit}}$, associated with minimal hysteresis-loss dispersion, is primarily governed by nanoparticle diameter and excitation frequency, while the influence of $\sigma_r$ remains secondary. 

The results displayed in Fig. \ref{Final_Fig_8} define an optimization scenario for MFH, which can be practically accessed by tuning the field amplitude. Remarkably, the predicted critical fields fall within the biological acceptable window for \textit{in vivo} biomedical applications, usually expressed in terms of the $H_{max}\cdot{f}$ constraint. 
Taking the reference value reported by the Atkinson and collaborators\cite{atkinson2007usable}, $f \cdot H_{max} \leq 4.8 \times 10^{8} \ \mathrm{A/(m \cdot s)}$ (often referred to as Brezovich limit\cite{brezovich1988low}), for $f=100$ kHz the corresponding maximum field amplitude is $\mu_0H_{max}\sim{6.1}$ mT, i.e. very close to the $H_{crit}$ values obtained for $D=30$ and 25 nm particles. 
It is worth noting that Pankhurst and collaborators\cite{kwok2023nonspecific} suggested that reducing the duty cycle would allow the safe threshold to be increased to approximately twice the Brezovich limit. In that case, again for $f=100$ kHz, the upper field limit would be $\mu_0H_{max}\sim{12.5}$ mT, which also lies within the range of critical fields shown in Fig.~\ref{Final_Fig_8}. Beyond biomedical applications, the results shown in Fig. \ref{Final_Fig_8} may also have significant implications for other non-biomedical fields. 
For example, they are directly relevant for sequential catalytic reactions requiring well-separated triggering temperatures\cite{ovejero2021selective}. In such cases, the absence of physiological constraints on $f \cdot H_{\max}$ would in principle allow independent tuning of the frequency to control the target SLP, while still operating within a regime of minimized local-heating dispersion.

\begin{figure}[H]
    \centering
    \includegraphics[width=1.0\linewidth]{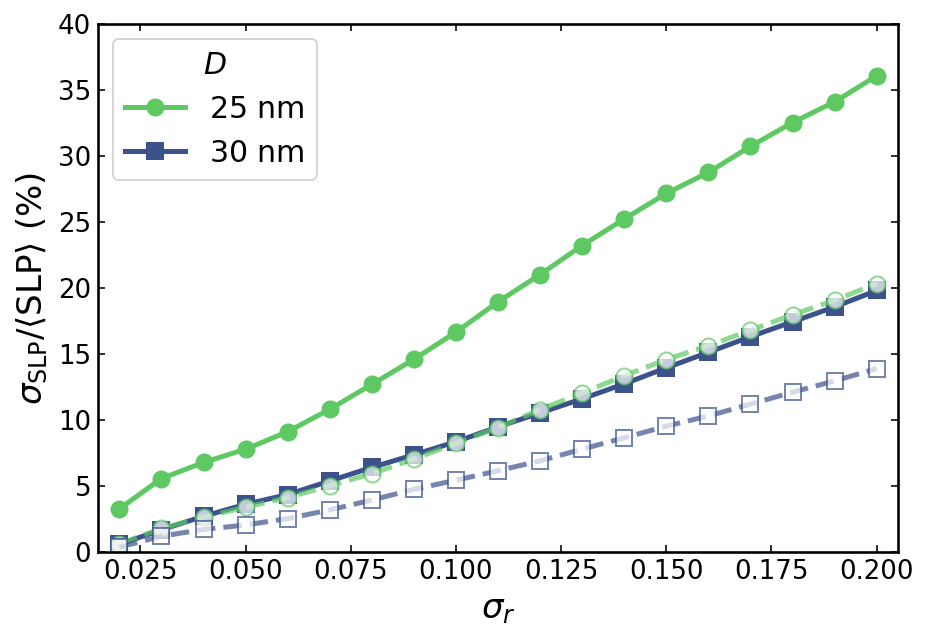}
    \caption{Variation of $\sigma_{\mathrm{SLP}} / \langle \mathrm{SLP} \rangle$ as a function of $\sigma_r$, for $D=30$ nm (blue squares) and $D=25$ nm (orange circles), respectively. Empty symbols (dashed lines) correspond to $f=1000$~kHz, and filled symbols (solid lines) to $f=100$~kHz.}
    \label{Final_Fig_9}
\end{figure}

It is important to emphasize that although Fig. ~\ref{Final_Fig_8} identifies an optimal field amplitude for given $D$ and $f$ conditions that appears to be only weakly dependent on $\sigma_r$, a finite dispersion in heat production persists and must be taken into account for any practical application.
%while the values in Fig. \ref{Final_Fig_8} indicate an optimising field amplitude for given $D$ and $f$ conditions that seems roughly independent on the specific $\sigma_r$ value, still there is an associated dispersion in heat production that needs to be taken into account for any possible application. 
To quantify its magnitude, Fig. \ref{Final_Fig_9} shows the corresponding values of $\sigma_{\mathrm{SLP}} / \langle \mathrm{SLP} \rangle$ associated with the data presented in Fig.~\ref{Final_Fig_8}.
 In Fig. \ref{Final_Fig_9} is clearly observed that while the position of $H_{\mathrm{crit}}$ is only marginally affected by shape polydispersity, there is a clear and systematic increase in the relative dispersion of the SLP with increasing $\sigma_r$, for the different particle sizes and excitation frequencies. This behaviour demonstrates that shape-induced polydispersity substantially enhances local heating fluctuations, even at the field at which hysteresis losses are minimized. Consequently, more polydisperse systems exhibit more pronounced spatial variations in heat dissipation, whereas highly monodisperse ensembles show markedly reduced local losses at $H_{\mathrm{crit}}$. These results underscore the importance of achieving high monodispersity to ensure spatially homogeneous and efficient heating in MFH applications.
% Added by Oscar 22 Jan 18:35
 These trends further indicate that, while field tuning provides an effective means to reduce local heating variability, it cannot fully compensate for strong shape disorder. Therefore, achieving highly homogeneous heating in MFH requires a combined strategy involving both optimized excitation conditions and careful control over nanoparticle shape distributions during synthesis.

% Fig~\ref{Final_Fig_9} shows the dependence of the 
% $\sigma_{\mathrm{SLP}} / \langle \mathrm{SLP} \rangle \%$ values corresponding to $H_{\mathrm{crit}}$ on $\sigma_r$. As discussed in Fig~\ref{Final_Fig_8}, although  $H_{\text{critic}}$ is not strongly sensitive to $\sigma_r$, the SLP values are significantly influenced by the $\sigma_r$ parameter. As the polydispersity of shape-induced anisotropy increases, the variations in the normalized $\sigma_{\mathrm{SLP}} / \langle \mathrm{SLP} \rangle$ values become more pronounced for small nanoparticles and at low frequencies, while larger particles and higher frequencies show comparatively smaller variations, indicating that the heating efficiency in hyperthermia treatment can become irregular, particularly for small nanoparticles.

\section{Conclusions}

We have presented a detailed numerical study of the interplay between shape‑induced anisotropy polydispersity and applied magnetic field amplitude on both global and local heating performance in magnetic fluid hyperthermia. By employing real‑time magnetization dynamics simulations and explicitly accounting for the coexistence of intrinsic cubic magnetocrystalline anisotropy and shape‑induced uniaxial anisotropy in magnetite nanoparticles, we extend previous theoretical descriptions\cite{munoz2017towards} toward a more realistic representation of experimental systems.

A central result of this study is the demonstration that ensemble‑averaged heating efficiency,  $\langle \mathrm{SLP} \rangle$, and local heating homogeneity, reported in terms of standard deviation of SLP, $\sigma_{\mathrm{SLP}}$ are not trivially correlated. While the global specific loss power increases monotonically with increasing field amplitude, the dispersion of single‑particle heat losses shows a non‑monotonic behavior when evaluated relative to the ensemble average. In particular, the analysis of the normalized standard deviation, $\sigma_{\mathrm{SLP}} / \langle \mathrm{SLP} \rangle$, reveals a local critical field, $H_{\mathrm{crit}}$, at which local hysteresis losses are minimized relative to global losses. This confirms, using a more comprehensive physical model and real‑time dynamics, the existence of an optimal operating field previously predicted\cite{munoz2017towards}.

The range of $H_{\mathrm{crit}}$, is governed primarily by particle diameter and field frequency, while 
the influence of polydispersity ($\sigma_r$) remains minor. A well-defined $H_{\mathrm{crit}}$ 
is observed for larger particles ($D = 25-30$ nm), whereas smaller ($D = 15-20$ nm) or monodisperse particles do not exhibit such behavior. These results underscore the key roles of particle size, shape-anisotropy distribution, and field frequency in achieving uniform and efficient heating  in magnetic hyperthermia. 

Regarding the direct translation of these theoretical results to practical implementations, it is important to keep in mind some key factors.
First, the results were specifically obtained for magnetite nanoparticles with an average aspect ratio dispersion of $<r>=1.1$. Other compositions and/or elongations would give very different uniaxial contributions\cite{usov2012magnetic}. Second, we considered a size-monodisperse system, while it is known that size polydispersity is intrinsic to a real system, and has a strong influence on local (single-particle) heating\cite{munoz2016distinguishing}. Third, the simulations were run for non-interacting conditions, which would need to be taken into account if aggregates are formed\cite{usov2019heating,russier2013}. 

Another aspect that deserves special attention is that, while the easy axes were assumed to be randomly oriented, the average heating performance was computed for a given $r$ category without considering the actual particle elongation direction relative to the applied field.
However, the orientation of the anisotropy axes with respect to the field can have a significant impact on heating performance\cite{conde2015orientation}. 
Moreover, the present treatment implicitly corresponds to
%the latter consideration lays within the \
a textit{frozen ferrofluid} scenario. 
In contrast, in a viscous environment where particles are free to rotate, a dynamical reorientation of the anisotropy axes may occur\cite{usov2010low,ota2017evaluation}, leading to additional modifications of the heating performance\cite{mamiya2011hyperthermic,simeonidis2016situ}.

Taken together, these considerations indicate that identifying optimal $H_{crit}$ conditions for a given nanoparticle sample may require a dedicated analysis tailored to the specific physical conditions.
%a specific study to seek possible $H_{crit}$ conditions would be required for a particular sample of magnetic nanoparticles to be used, at least based on the previous results. 
While the present results provide general guidelines for minimizing heating heterogeneity, extending them to more complex and realistic scenarios involving orientational effects and particle mobility will require further investigation.
%Extensive further work is required to understand the optimising field conditions for MFH obtained here to such a far more complex scenario.

\section*{Author contributions}

Conceptualization: D.S. Data curation: N. \c{C}., I.L.V.
Formal analysis: N. \c{C}., I.L.V., D.S., Ò.I Funding acquisition: D.S. Investigation: N. \c{C}., I.L.V., D.S., Ò.I
Methodology: N. \c{C}., I.L.V., D.S., Ò.I
Project administration: D.S., Ò.I
Software: N. \c{C}., I.L.V.
Supervision: D.S., Ò.I
Validation: N. \c{C}., I.L.V. , D.S., Ò.I
Visualization: N. \c{C}., I.L.V.
Writing – original draft: N. \c{C}.
Writing – review \& editing: N. \c{C}., I.L.V., D.S., Ò.I.

\section*{Conflicts of interest}
There are no conflicts to declare.

\section*{Data availability}

All the simulation results presented in this article have been obtained using the micromagnetic simulator OOMMF, available from the NIST website, https://math.nist.gov/oommf/software.html. 
%\textcolor{red}{
%Oscar: we could upload all the result files and scripts to produce the Figs. to a digital repository such as Zenodo or similar (!!) Some articles do it, but it involves organizing and classifying into folders everything.
%}

\section*{Acknowledgements}

We acknowledge financial support by Spanish Ministerio de Ciencia, Innovaci\'on y Universidades through projects PID2024-157172NB-I00 and CNS2024-154574, "ERDF A way of making Europe", by the "European Union", and Catalan DURSI (2021SGR0032). Xunta de Galicia is acknowledged for projects ED431F 2022/005 and ED431B 2023/055. AEI is also acknowledged for  the \textit{Ram\'on y Cajal} grant RYC2020-029822-I that supports the work of D.S. We acknowledge the Centro de Supercomputacion de Galicia (CESGA) for computational resources.

%%%END OF MAIN TEXT%%%


\balance


%%%REFERENCES%%%
\bibliography{rsc} %You need to replace "rsc" on this line with the name of your .bib file
\bibliographystyle{rsc} %the RSC's .bst file
\end{document}